\documentclass[journal]{IEEEtran}
\usepackage{amsmath,amsfonts}
\usepackage[caption=false,font=normalsize,labelfont=sf,textfont=sf]{subfig}
\usepackage{stfloats}
\usepackage{url}
\usepackage{graphicx}
\usepackage{cite}
\usepackage{physics}
\usepackage{mleftright}
\usepackage{hyperref}

\DeclareMathOperator{\diag}{diag}

\begin{document}

\title{Polarization-Dependent Loss Mitigation via Orthogonal Design Precoding and Interference Cancellation}

\author{Mohannad Shehadeh,~\IEEEmembership{Graduate Student Member,~IEEE} and Frank R.~Kschischang,~\IEEEmembership{Fellow,~IEEE}
\thanks{The authors are with the Edward S. Rogers Sr. Department of
	Electrical \& Computer Engineering, University of Toronto, Toronto, ON M5S 3G4,
	Canada. Emails: \protect\scalebox{0.9}[1.0]{\texttt{\{mshehadeh,frank\}@ece.utoronto.ca.}}}
\thanks{Submitted for publication on Feb.~6, 2025. Resubmitted on Mar.~14, 2025.}}

\markboth{IEEE Photonics Technology Letters}{Shehadeh and Kschischang: Polarization-Dependent Loss Mitigation with Orthogonal-Design-Based Precoding and Interference Cancellation}

\IEEEpubid{0000--0000/00\$00.00~\copyright~2025 IEEE}

\maketitle


\begin{abstract}
Recent work by Shehadeh and Kschischang provides 
a simple capacity-achieving scheme
for channels with 
polarization-dependent loss (PDL) under common modeling 
assumptions via a careful choice of
orthogonal-design-based precoding and interference cancellation.
This letter extends that
work with a simulation-based
demonstration showing that this 
scheme remains highly effective
at mitigating PDL in the
highly practical setting of 
16-QAM with Chase-decoded extended Hamming 
inner codes rather than
the near-capacity inner codes considered 
in the original work. 
An alternative near-optimal variation of this scheme is also provided
requiring only one inner code rather 
than two and suffering no penalty in the 
absence of PDL, making it much more practical.
\end{abstract}


\section{Setup}\label{setup-section}
\IEEEPARstart{W}{e} consider a memoryless 
model for a polarization-division-multiplexed (PDM)
coherent optical communication system with 
slowly-varying polarization-dependent loss (PDL)
and deterministic insertion loss. 
We further assume that the channel parameters 
are perfectly 
known to the receiver but unknown
to the transmitter. Such a model is considered in 
	\cite{jlt-pdl-me,dumenil-dissertation,dumenil-jlt,
	huawei,damen,awwad-dissertation,old-pt,
	winzer-shtaif,oyama}
and it is shown by Shehadeh and Kschischang
in \cite{jlt-pdl-me} that a certain
orthogonal-design-based precoding and interference 
cancellation scheme is capacity-achieving under this model.

By a standard reduction technique, the channel matrix
under this model is given by
\begin{equation*}
		\underbrace{\begin{pmatrix}
			\sqrt{1+\gamma} &  0\\
			 0 & \sqrt{1-\gamma}	
	\end{pmatrix}}_{\mathbf{D}_\gamma}
	\underbrace{\begin{pmatrix}
			\cos\theta & -\sin\theta \\
			\sin\theta & \cos\theta
	\end{pmatrix}}_{\mathbf{R}_\theta}
	\underbrace{\begin{pmatrix}
			e^{i\phi} & 0\\
			0 & e^{-i\phi}	
	\end{pmatrix}}_{\mathbf{B}_\phi}
\end{equation*}
which yields a three-parameter channel class
\begin{equation*}
	\mathbf{Y}
	=
	\mathbf{D}_\gamma \mathbf{R}_\theta \mathbf{B}_\phi
	\mathbf{X}
	+ 
	\mathbf{Z}
\end{equation*}
where  $\gamma \in [-\alpha,\alpha]$, $\theta \in [0,2\pi)$,
and $\phi \in [0,2\pi)$. This
represents an adversarial \emph{class} of channels corresponding
to two polarizations with up to 
\begin{equation*}
	10\log_{10}\mleft(\frac{1+\alpha}{1-\alpha}\mright)\text{ dB}
\end{equation*}
of PDL where $\alpha \in [0,1)$ is a fixed parameter.
Moreover, $\mathbf{X}$ and $\mathbf{Z}$ are independent
with $\mathbf{Z}$ being standard white Gaussian 
and $\mathbf{X}$ satisfying
$\mathbb{E}[\norm{\mathbf X}_2^2] 
	= 2\cdot \mathsf{SNR}$ where $\mathsf{SNR}$
	denotes the signal-to-noise ratio (SNR). We are
generally interested in the \emph{worst-case}
performance of any given scheme over all 
$\gamma \in [-\alpha,\alpha]$, $\theta \in [0,2\pi)$,
and $\phi \in [0,2\pi)$.
	
A common simplification whose justification is discussed
in \cite{dumenil-dissertation} is to take 
$\phi = 0$ and thus remove $\mathbf{B}_\phi$ from the model.
In this paper,
we will cover \emph{both} the simplified case
of $\phi = 0$ \emph{and} the general case 
of adversarial $\phi \in [0,2\pi)$. In the case
of $\phi = 0$, it suffices to precode only
across in-phase (I) and quadrature (Q) components and
not across time, while in the general case of 
$\phi \in [0,2\pi)$, we must also precode across
time. We will initially consider the case of 
$\phi = 0$ for the
purposes of exposition and because it is of interest
in its own right, with the case of $\phi \in [0,2\pi)$
deferred to the end. 
\IEEEpubidadjcol

We will work with real-valued
\emph{equivalent representations} of \emph{complex-valued}
vectors and matrices where the first and second halves of
the \emph{real-equivalent} of 
a vector contain its real and imaginary parts 
respectively. This simplifies the description
of I/Q precoding and the necessary 
widely-linear processing to \emph{equivalent} 
linear
processing on the concatenated real--imaginary
representations. When $\phi = 0$, 
the channel matrix is already real-valued
and its real-equivalent representation is
thus $\diag(\mathbf{D}_\gamma\mathbf{R}_\theta, \mathbf{D}_\gamma\mathbf{R}_\theta)\in 
\mathbb{R}^{4\times 4}$. 
We take the input to
the channel to 
be $\mathbf{X} = \mathbf{G}\mathbf{U}$ where $\mathbf{X},\mathbf{U} \in 
\mathbb{R}^{4\times 1}$ and $\mathbf{G} \in \mathbb{R}^{4\times 4}$
is an \emph{orthogonal} precoding matrix. 
The \emph{effective channel} is then described
by the $4 \times 4$ effective channel matrix 
$\diag(\mathbf{D}_\gamma\mathbf{R}_\theta, \mathbf{D}_\gamma\mathbf{R}_\theta)\mathbf{G}$
and has input $\mathbf{U}$ satisfying 
$\mathbb{E}[\norm{\mathbf U}_2^2] 
= 4\cdot \mathsf{SNR}$.

We use bit-interleaved coded modulation
(BICM) \cite{BICM} with extended Hamming codes 
and a $16$-QAM constellation. This corresponds to $\{-3,-1,1,3\}$-signaling \emph{in each degree of freedom} under our real-equivalent
representation.
We further use soft-decision Chase decoding \cite{Chase}
in conjunction with all schemes considered.
Since the scheme of \cite{jlt-pdl-me} requires
two codes with different rates used in equal proportion, 
we choose them to have the same
average rate as a third code which is used in 
schemes requiring one code. In particular,
we consider three 
codes with lengths and dimensions $(n_1,k_1)$, 
$(n_2,k_2)$, and $(n,k)$ respectively satisfying
$$
\frac{1}{2}\left(\frac{k_1}{n_1} + \frac{k_2}{n_2}\right) = \frac{k}{n}\text{.}
$$
We take $(n_1,k_1) = (96,86)$, 
$(n_2,k_2) = (384,372)$, and $(n,k)=(192,179)$.
These codes are obtained by shortening longer extended
Hamming parent codes.
This yields a rate of 
approximately $0.93$ so that the inner
coded modulation scheme altogether is somewhat comparable 
to that in the 400ZR 
implementation agreement \cite{400ZR}.

Our key contributions are the demonstration 
of the effectiveness 
of the scheme of \cite{jlt-pdl-me} for
PDL mitigation in this context and 
the introduction of a variation on this scheme 
with various practical advantages. We illustrate
our points by comparing five schemes defined in
Section \ref{five-schemes-section} which include
the scheme of \cite{jlt-pdl-me} and the proposed
variation followed by simulations 
in Section \ref{simulation-section}. 
We then address
more general scenarios in Section \ref{Further-Results-Section}
before concluding in
Section \ref{conclusion-section}.

\section{Five Schemes}\label{five-schemes-section}

We now describe five schemes which we
identify as iZ, pZ, D, pD, and the scheme of \cite{jlt-pdl-me}.
These are based on precoding, linear minimum mean 
square error (LMMSE) equalization, and 
successive interference cancellation (SIC) \cite[Chapter~8]{tse-wireless-communication}.
We use zero-forcing (ZF) instead of LMMSE
for simplicity since it is comparable at the high SNRs of interest. We will assume 
$\phi = 0$ and I/Q precoding as described in 
Section \ref{setup-section} for the purposes
of describing our five schemes in this section. 
Implementation details 
for all schemes in both the $\phi = 0$ case and
the general $\phi \in [0,2\pi)$ case can be found
in our complete simulation code which we make available online \cite{pdl-scheme-sim}.

\subsection{One Code, Spatio-Temporal Interleaving, and ZF without Precoding (iZ)}

Scheme iZ is our baseline scheme in which 
we interleave two codewords from \emph{one} code 
across the
two polarizations \emph{and} in time to decorrelate the noise.
We then equalize via ZF and decode each codeword independently. 
This is easy to do and achieves some PDL penalty mitigation
relative to treating each polarization completely independently.

\subsection{One Code and ZF with Precoding (pZ)}

In Scheme pZ, we again send two codewords from \emph{one} code but 
consider a two-channel-use extension and the $4\times 4$
precoding matrix from \cite{jlt-pdl-me}:
\begin{equation}\label{precoder}
	\mathbf{G} = \frac{1}{\sqrt{2}}
	\begin{pmatrix}
		1 & 0 & 1 & 0\\
		0 & 1 & 0 & 1\\
		0 & 1 & 0 & -1\\
		-1 & 0 & 1 & 0
	\end{pmatrix}\text{.}
\end{equation} This is followed
by ZF and independent
decoding of each codeword which sees 
a worst-case effective gain of $1-\alpha^2$. 
The resulting scheme is essentially equivalent
to those of \cite{pairwise,oyama}
as shown in \cite{jlt-pdl-me}. In particular,
I/Q and polarization--time coding are 
performance-equivalent when $\phi = 0$.

\subsection{One Code and D-BLAST-Style ZF-SIC without Precoding (D)}

In Scheme D, we consider a \emph{D-BLAST-style} \cite[Chapter~8]{tse-wireless-communication} approach 
which allows us to perform ZF-SIC on one codeword by diagonally
staggering it in time and space. After Gray-mapping to
$\{-3,-1,1,3\}$, we have a codeword of $n/2$ symbols which is
split into two halves $\mathbf x^\mathsf{f}$ and $\mathbf x^\mathsf{s}$
of $n/4$ symbols. We then transmit $\zeta$ such codewords in 
$(\zeta+1)n/4$ channel uses in time which we represent by the
$2 \times (\zeta+1)n/4$ matrix 
$$
\begin{pmatrix}
	\psi & \mathbf x_1^\mathsf{s} & \mathbf x_2^\mathsf{s} & \cdots & \mathbf x_{\zeta-2}^\mathsf{s} & \mathbf x_{\zeta-1}^\mathsf{s} & \mathbf x_\zeta^\mathsf{s}\\
	\mathbf x_1^\mathsf{f} & \mathbf x_2^\mathsf{f} & \mathbf x_3^\mathsf{f}  & \cdots & \mathbf x_{\zeta-1}^\mathsf{f} & \mathbf x_\zeta^\mathsf{f} & \psi
\end{pmatrix}
$$
where $\psi$ denotes some fixed symbol sequence known at the receiver and transmitter.
This results in an overall rate of $$
\frac{\zeta}{\zeta+1} \cdot \frac{k}{n}
$$
with $\zeta$ controlling the rate loss at the cost of increased
impact of error propagation. We then 
decode $\zeta$ codewords by repeating the following
steps for each column in the above:
\begin{itemize}
	\item If this is the first column, cancel the 
	interference corresponding to $\psi$; otherwise,
	cancel the interference corresponding to the 
	second half estimate obtained in the previous step.
	Equalize after interference cancellation to produce a noisy first half for the next step.
	\item 
	Equalize the next column, append the second half of our noisy codeword to the noisy first half obtained in the previous step, and decode it. We use the
	second half of the decoded codeword as the estimate for the next step.
\end{itemize}
What this accomplishes relative to iZ is that the first
half of \emph{every} codeword always sees a cleaner interference-free 
channel. However, this \emph{usually} cannot improve upon 
iZ in this setting since the case of 
$\theta = 0$, which is already interference-free, 
is \emph{usually} the worst case. 
In particular, for both iZ and D, 
when $\theta = 0$ and $\gamma = \pm\alpha$, the two halves
of each codeword see effective gains of 
$1+\alpha$ and $1-\alpha$. This is usually the worst case 
for off-the-shelf coded modulation schemes which are
not designed for gain imbalance.  
This leads to the next scheme.

\begin{figure}[!t]
	\centering
	\includegraphics[width=\columnwidth]{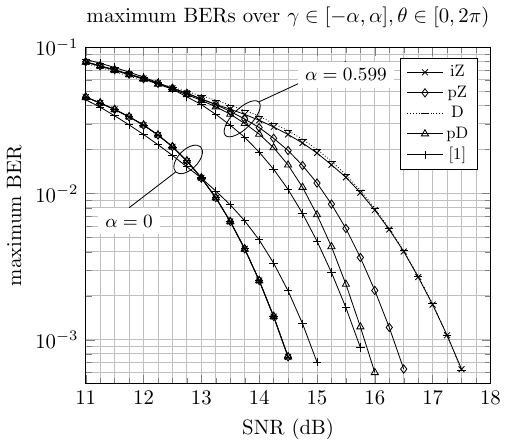}
	\caption{Simulation results for Schemes iZ, pZ, D, pD, and
the scheme of \cite{jlt-pdl-me} for $\alpha = 0$ and $\alpha = 0.599$ corresponding respectively
		to worst-case PDLs of $0$ dB and $6$ dB.}
	\vspace{-10pt}
	\label{sims-fig}
\end{figure}

\subsection{One Code and D-Blast-Style ZF-SIC with Precoding (pD)}

In Scheme pD, we consider combining the orthogonal-design-based
precoding and interference cancellation approach
of \cite{jlt-pdl-me} with a D-BLAST-style
technique as in Scheme D so that only a single code is used. 
In particular, 
we consider a two-channel-use extension 
and the
precoding matrix \eqref{precoder}
after which we apply Scheme D to the resulting 
effective channel. After Gray-mapping to
$\{-3,-1,1,3\}$, 
we have a codeword of $n/2$ symbols which we
split into two halves $\mathbf u^\mathsf{f}$ and $\mathbf u^\mathsf{s}$
of dimensions $2 \times n/8$ containing $n/4$ symbols each. 
We then transmit $\zeta$ such codewords in 
$(\zeta+1)n/8$ uses of the two-channel-use extended channel 
so that the input is represented by the 
$4 \times (\zeta+1)n/8$ matrix 
$$
\mathbf{G}
\begin{pmatrix}
	\psi & \mathbf u_1^\mathsf{s} & \mathbf u_2^\mathsf{s} & \cdots & \mathbf u_{\zeta-2}^\mathsf{s} & \mathbf u_{\zeta-1}^\mathsf{s} & \mathbf u_\zeta^\mathsf{s}\\
	\mathbf u_1^\mathsf{f} & \mathbf u_2^\mathsf{f} &  \mathbf u_3^\mathsf{f} & \cdots & \mathbf u_{\zeta-1}^\mathsf{f} & \mathbf u_\zeta^\mathsf{f} & \psi
\end{pmatrix}
$$
where $\psi$ is some fixed $2 \times n/8$ symbol sequence.
As before, we have an overall rate of $\zeta/(\zeta+1) \cdot k/n$.
We then decode exactly as in Scheme D but we consider the
effective channel combined with $\mathbf{G}$ for the purposes
of equalization and interference cancellation. 

Unlike the case of
Scheme D, the two halves
of each codeword respectively see effective gains of 
$1$ and $1-\alpha^2$. This is due to the fact 
that \eqref{precoder} induces an \emph{orthogonal design} \cite{orthogonal-designs-book,orthogonal-designs-paper} 
in each half of the effective channel matrix as shown in \cite{jlt-pdl-me}. 
While we expect a gain
over pZ since half of our symbols have a
strictly improved reliability, this scheme will 
still be usually sub-optimal due to the remaining (but reduced)
gain imbalance.

\subsection{Two Codes and ZF-SIC with Precoding ([1])}

Lastly, we consider the
scheme proposed in \cite{jlt-pdl-me}
using two codes with average rate $k/n$ as described 
in Section \ref{setup-section}. This entails precoding as 
in pZ but sending two codewords from the 
\emph{two} different
codes \emph{and} using ZF-SIC rather than ZF.
Due to the orthogonal design associated with 
\eqref{precoder},
the lower rate and thus stronger code always 
sees a worst-case gain of 
$1-\alpha^2$ while the higher rate and thus weaker 
code always sees a worst-case gain of $1$. 
As shown in \cite{jlt-pdl-me}, the resulting
scheme is strictly optimal
in the sense that the performance is only 
limited by the classical additive white 
Gaussian noise channel performance of 
the two codes used and the fundamental 
information-theoretic cost of PDL.

\section{Simulations}\label{simulation-section}

\begin{figure}[!t]
	\centering
	\includegraphics[width=\columnwidth]{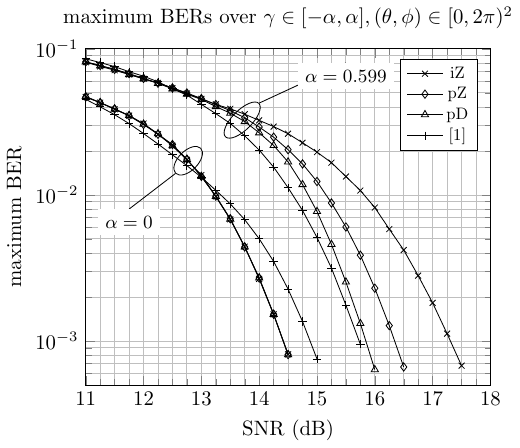}
	\caption{Simulation results for Schemes iZ, pZ, pD, and
		the scheme of \cite{jlt-pdl-me} for $\alpha = 0$ and $\alpha = 0.599$ corresponding respectively
		to worst-case PDLs of $0$ dB and $6$ dB
		and in the case of a general strictly complex-valued channel.}\vspace{-10pt}
	\label{sims-fig-complex}
\end{figure}

We now provide bit error rate (BER) versus
SNR simulation results for each of the five
schemes.
For Schemes D and pD, we take $\zeta  = 100$. Chase 
decoding parameters and other details can be found
in our simulation code \cite{pdl-scheme-sim}. 
We plot in Fig.~\ref{sims-fig} the worst-case (maximum) BER over all
$\gamma\in[-\alpha,\alpha]$ and $\theta\in[0,2\pi)$
(which we discretize) for $\alpha = 0$ and $\alpha = 0.599$
corresponding respectively to channel classes with a
worst-case PDL of $0$ dB and a worst-case PDL of $6$ dB.

Fig.~\ref{sims-fig} shows that pD is
worse than the scheme of \cite{jlt-pdl-me} by about $0.25$ dB at a BER of 
$10^{-2}$ and by about $0.125$ dB at a BER of $10^{-3}$. 
However, the scheme of \cite{jlt-pdl-me} alone suffers
a significant penalty in the absence of PDL 
($\alpha = 0$) 
for 
BERs below $10^{-2}$ relative to all other schemes
considered. 
This is because the two different rates
of the two codes used must be tuned
according to the worst-case PDL value.
In practice, we may wish for a single system to
perform well both in the presence and the absence
of PDL and thus prefer pD. 
Moreover, pD eliminates the problem of
choosing two codes altogether making it trivial
to combine immediately with any single off-the-shelf
coded modulation scheme.

Fig.~\ref{sims-fig} also highlights some subtle points discussed: 
While iZ and D are almost indistinguishable 
as anticipated,
D is very slightly, but measurably, better than iZ
at very low SNRs 
since $\theta = 0$ is no longer the worst angle. In particular,
seeing $1+\alpha$ and $1-\alpha$ gains becomes 
better than seeing uniform $1-\alpha^2$ gains 
when $\theta = \pi/4$ since 
$1+\alpha$ acts as a diversity gain. Scheme D then
has an advantage in this new worst case. While not
particularly important here, this becomes relevant in
other contexts. For example, if a 
spatially-coupled
code with 
universality properties 
\cite{SC-Schmalen,spatially-coupled-are-universal} is used,
theory suggests that D will be near-optimal since 
such codes would be agnostic to the gain imbalance
as long as the mutual information is sufficient. If
realized, this would eliminate the need for precoding
and thus eliminate the peak-to-average power ratio
(PAPR) cost of precoding.

Lastly, we note that both pD and the scheme of \cite{jlt-pdl-me} suffer a 
performance loss in Fig.~\ref{sims-fig} 
due to error propagation in the interference cancellation.
This loss is virtually eliminated if interference cancellation is
done after decoding a hypothetical concatenated outer code
and the performance is considered at lower BERs. This is
also a practical possibility and increases the gains
of pD and the scheme of \cite{jlt-pdl-me} relative to pZ.

\section{Further Results}\label{Further-Results-Section}

When $\phi \in [0, 2\pi)$, the channel matrix 
is complex-valued and has $4 \times 4$ real-equivalent 
representation given by 
\begin{equation*}
	\mathbf{H}_{\gamma,\theta,\phi} =
	\begin{pmatrix}
		\mathbf{D}_\gamma & \mathbf{0}\\
		\mathbf{0} & \mathbf{D}_\gamma
	\end{pmatrix}
	\begin{pmatrix}
		\mathbf{R}_\theta & \mathbf{0}\\
		\mathbf{0} & \mathbf{R}_\theta
	\end{pmatrix}
	\begin{pmatrix}
		\Re (\mathbf{B}_\phi) & -\Im (\mathbf{B}_\phi)\\
		\Im (\mathbf{B}_\phi) & \Re (\mathbf{B}_\phi)
	\end{pmatrix}
\end{equation*}
which we extend to
$\diag(\mathbf{H}_{\gamma,\theta,\phi},\mathbf{H}_{\gamma,\theta,\phi})\in 
\mathbb{R}^{8\times 8}$ representing two channel uses in time.
We then take the input to 
be $\mathbf{X} = \mathbf{G}\mathbf{U}$ where $\mathbf{X},\mathbf{U} \in 
\mathbb{R}^{8\times 1}$ and $\mathbf{G} \in \mathbb{R}^{8\times 8}$ is our orthogonal precoding matrix. This yields
an effective $8 \times 8$ channel matrix given by 
$\diag(\mathbf{H}_{\gamma,\theta,\phi},\mathbf{H}_{\gamma,\theta,\phi})\mathbf{G}$ and having input 
$\mathbf{U}$ satisfying 
$\mathbb{E}[\norm{\mathbf U}_2^2] 
= 8\cdot \mathsf{SNR}$.
We then consider the $8 \times 8$ precoding
matrix from \cite{jlt-pdl-me}
\begin{equation}\label{precoder-complex}
	\mathbf{G}
	=
	\frac{1}{\sqrt{2}}
	\begin{pmatrix}
		1 & 0 & 0 & 0 & 1 & 0 & 0 & 0\\ 0 & 0 & 1 & 0 & 0 & 0 & 1 & 0\\ 0 & -1 & 0 & 0 & 0 & -1 & 0 & 0\\ 0 & 0 & 0 & -1 & 0 & 0 & 0 & -1\\ 0 & 0 & 1 & 0 & 0 & 0 & -1 & 0\\ -1 & 0 & 0 & 0 & 1 & 0 & 0 & 0\\ 0 & 0 & 0 & 1 & 0 & 0 & 0 & -1\\ 0 & -1 & 0 & 0 & 0 & 1 & 0 & 0 
	\end{pmatrix}
\end{equation}
and straightforwardly apply the schemes of Section \ref{five-schemes-section} with the effective 
channel split into two equal halves as before. The complete
implementation details are available in our 
complete simulation code \cite{pdl-scheme-sim}.
Simulation results are provided in Fig.~\ref{sims-fig-complex} and are
nearly identical to the case of $\phi = 0$
as predicted by \cite{jlt-pdl-me}. Scheme
D is omitted since it was merely pedagogical.

\begin{figure}[!t]
	\centering
	\includegraphics[width=\columnwidth]{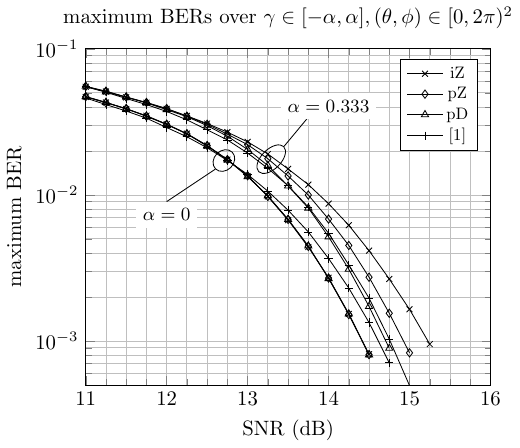}
	\caption{Simulation results for Schemes iZ, pZ, pD, and
		the scheme of \cite{jlt-pdl-me} for $\alpha = 0$ and $\alpha = 0.333$ corresponding respectively
		to worst-case PDLs of $0$ dB and $3$ dB
		and in the case of a general strictly complex-valued channel.}\vspace{-10pt}
	\label{sims-fig-complex-3dB}
\end{figure}

Lastly, we consider the case of $3$ dB
of worst-case PDL in Fig.~\ref{sims-fig-complex-3dB}.
In this case, we change the parameters
of the code pair used in the scheme
of \cite{jlt-pdl-me} to
$(n_1,k_1) = (144,131)$ and
$(n_2,k_2) = (288,275)$ which 
have the same average rate
as before but are closer to each other to account for the smaller
PDL. Note that the slightly worse
performance of the scheme of 
\cite{jlt-pdl-me} in this 
case relative to pD is \emph{not} a contradiction
of the information-theoretic optimality of the scheme of \cite{jlt-pdl-me}. 
This is a finite blocklength
effect due to the worse gap to capacity of
the lower rate Hamming code in the code pair
used. The scheme of \cite{jlt-pdl-me} is always
better if all inner codes used are at the same
average gap to capacity. However, 
it is typically harder to achieve 
the same gap to capacity at lower rates that are still greater than $1/2$. 
This further
highlights the significant benefits of the proposed scheme pD which obviates all
of these issues and does not need
to be adapted to the
worst-case PDL value.

\section{Concluding Remarks}\label{conclusion-section}

We have illustrated that the scheme of \cite{jlt-pdl-me}
remains effective in practical coded modulation settings
and provided a variation (pD) which eliminates
the requirement of tuning two codes. Future work can
consider alternative approaches such as exploiting
the universality properties of spatially-coupled
codes \cite{SC-Schmalen} to eliminate the need for
precoding altogether.

\section*{Acknowledgments}
The authors would like to acknowledge Dr.\ Hamid Ebrahimzad 
for suggesting a variation of \cite{jlt-pdl-me} similar to 
pD and pointing out its practical benefits.


\end{document}